\documentclass{article}
\usepackage{amssymb}

\usepackage{graphicx}
\usepackage{amsmath}

\input{tcilatex}

\begin{document}

\title{\textbf{Duffin-Kemmer-Petiau and Klein-Gordon-Fock Equations for
Electromagnetic, Yang-Mills and external Gravitational Field Interactions:
proof of equivalence.}}
\author{V. Ya. Fainberg\thanks{%
Permanent address: P.N. Lebedev Institute of Physics, Moscow, Russia, e-mail
fainberg@lpi.ru} \ and B. M. Pimentel \\
Instituto de F\'{i}sica Te\'{o}rica\\
Universidade Estadual Paulista\\
Rua Pamplona, 145\\
01405-900-S\~{a}o Paulo\\
SP-Brasil}
\maketitle

\begin{abstract}
Starting from Generating functional for Green Function (GF), constracted
from Lagrangian action in Duffin-Kemmer-Petiau (DKP) theory ($\mathcal{L}$%
-approach) we strictly prove that the physical matrix elements of S-matrix
in DKP and Klein-Gordon-Fock (KGF) theories coincide in cases of interaction
spin O particles with external and quantized Maxwell and Yang-Mills fields
and in case of external gravitational field (without or with torsion). For
the proof we use reduction formulas of Lehmann Symanzik , Zimmermann (LSZ).
We prove that many photons and Yang-Mills particles GF coincide in the both
theories, too.
\end{abstract}

\section{Introduction}

The question on equivalence of DKP and KGF equations has a long history. In
1926 the second order relativistic equation for spin $0$ particle has been
discovered by O. Klein \cite{Klein}, V. Fock \cite{Fock}, W Gordon \cite
{Gordon} and others\footnote{%
See [4] where a rich list of references with interesting historical comments
can be found, specially about equivalence the DKP and KGF equations. For
references see also review [5], where in particular there is references to
works I. Gelfand and A.Yaglom, who obtained the firs order equation for
particle with aritrary spin ,too.}. After the appearance in 1928 of the
famous Dirac equation \cite{Dirac}, G. Petiau \cite{Petiau}, R. Duffin \cite
{Duffin} and N. Kemmer \cite{Kemmer} proposed in 1936-39 years the first
order DKP equation for description of spin $0$ and $1$ particles. From the
beginning of this time to prove equivalence between DKP and KGF equations
many attempts have been undertaken (see reference in \cite{Krajcik}). From
many questions connected with attempts to proof the equivalence of the both
theories we would like to discuss as minimum three important ones: 1. Is
there a strict proof of equivalence of the both theories in case of the
interaction charged spin $0$ particles with eletromagnetic field introduced
by minimal way? 2. For which types of interactions is it possible to give
strict proof of the equivalence of these both theories? 3. Is there
equivalence between these theories on description of process of decay
unstable particles (specially decay of $K_{L}^{0}$ mesons)? For details see
references in \cite{Krajcik}.

In QED many processes have been investigated in frame work of DKP theories
(see works \cite{Umezawa}-\cite{Pimentel} and others references in \cite
{Krajcik}). The main conclusion is that calculation based on DKP and KGF
equations yield identical results including one-loop corrections. However
the strict proof of equivalence of the both theories for physical matrix
elements of $S$-matrix does not exist even in QED for any order of the
perturbation theory. There was also no attempts to generalize the proof on
the other types of interactions. The question about the equivalence of the
both theories for description unstable particles is remained open.

The main goal of this paper is to strictly prove that matrix elements of $S$%
-matrix in DKP and KGF theories coincide in cases interaction spin $0$
particles with external and quantized Maxwell and Yang-Mills fields and also
in case of external gravitational fields (without and with torsion). For the
proof we utilized reduction formulas of Lehmann, Symanzik, Zimmermann (LSZ), 
\cite{Lehmann}, and Lagrangian approach to construction the generating
functional of GF in all the cases of interactions.

As concerned to description of decay unstable particles, we believe that the
both theories have to give the same results as well as for all process which
are formulated in terms of renormalizable theories. Although this question
goes beyond the scope of the paper we give in conclusion the simple example
of Lagranian for description of the decay $K_{L}^{0}$ meson in DKP theory.

The canonical Hamiltonian approach to quantization of DKP equations has been
developed in work \cite{Fainberg} where it has been proved that generating
functional for GF in DKP theory coincides with that of in Lagrangian
approach. Thus we start in this paper from generating functional for GF
constracted from Lagrangian action with external sources in exponent of
functional integral (named $\mathcal{L}$-approach). This approach makes
simpler the proof of the equivalence of the both theories .

In section 2 the generating functionals are constructed in DKP theory for GF
of spin $0$ particles interacted with Maxwell and Yang-Mills fields and with
external gravitational field (without torsion and with one). We also prove
that all many photons GF (not only matrix elements of $S$-matrix) coincide
on DKP and KGF theories.

In section 3 the equivalence of the matrix elements of $S$-matrix in the
both theories are proved for all above mentioned interactions, including
non-abelian theories. For strict proof of the results we utilized the wave
packets instead of plane waves. In section 4 are given short conclusions
about basic results.

\section{Generating functional}

First one apply $\mathcal{L}$-approach for construction of the generating
functional of GF in DKP theory for charged $0$-spin particles interacting
with quantized EM field $A_{\mu }(x)$. From general consideration the
generating functional in $\mathcal{L}$-approach has the following form (in $%
\alpha $-gauge)\footnote{%
Notation see in \cite{Fainberg}}: 
\begin{eqnarray}
Z\left( \overline{\mathcal{J}},\mathcal{J},\mathcal{J}_{\mu }\right)
&=&Z_{0}^{-1}\int \mathcal{D}A_{\mu }\mathcal{D}\psi \mathcal{D}\overline{%
\psi }\exp \left\{ i\int d^{4}x\left( -\frac{1}{4}F_{\mu \nu }F^{\mu \nu }-%
\frac{1}{2\alpha }\left( \partial _{\mu }A^{\mu }\right) ^{2}\right. \right.
\notag \\
&&\left. \left. +\mathcal{J}_{\mu }A^{\mu }+\overline{\psi }(x)\left( i\beta
_{\mu }\mathcal{D}^{\mu }-m\right) \psi (x)+\overline{\mathcal{J}}\psi +%
\overline{\psi }\mathcal{J}\right) \right\}  \label{ec1}
\end{eqnarray}
where 
\begin{equation}
Z_{0}=Z(0,0,0);\quad D^{\mu }=\partial ^{\mu }-ieA^{\mu }  \label{ec2}
\end{equation}

After integration over $\psi \left( x\right) $ and $\overline{\psi }\left(
x\right) $ we get: 
\begin{eqnarray}
Z\left( \overline{\mathcal{J}},\mathcal{J},\mathcal{J}_{\mu }\right)
&=&Z_{0}^{-1}\int \mathcal{D}A_{\mu }\exp \left\{ i\int d^{4}x\left( -\frac{1%
}{4}F_{\mu \nu }F^{\mu \nu }-\frac{1}{2\alpha }\left( \partial _{\mu }A^{\mu
}\right) ^{2}\right. \right.  \notag \\
&&\left. \left. +Tr\ln S\left( x,x,A\right) -\int d^{4}y\overline{\mathcal{J}%
}\left( x\right) S\left( x,y,A\right) \mathcal{J}\left( y\right) \right)
\right\}  \label{ec3}
\end{eqnarray}
here 
\begin{equation}
S\left( x,y,A\right) =\left( i\beta _{\mu }D^{\mu }-m\right) ^{-1}\delta
^{4}\left( x-y\right)  \label{ec4}
\end{equation}
is GF of particles in the field $A_{\mu }\left( x\right) $; the term $Tr\ln
S\left( x,x,A\right) $ is responsible for arising of the all diagrams of
vacuum polarization.This term can be transformed to the following view%
\footnote{%
We utilized the equation (\ref{ec26}), which expresses $\psi _{\alpha }(x),%
\overline{\psi }_{\alpha }(x)$ through the components.}, 
\begin{eqnarray}
\det S\left( x,y,A\right) &=&C\int \mathcal{D}\psi \mathcal{D}\overline{\psi 
}\exp \left\{ i\int \overline{\psi }\left( i\beta _{\mu }D^{\mu }-m\right)
\psi \right\}  \label{ec5} \\
&=&C\int \underset{\alpha }{\Pi }\mathcal{D}\varphi _{\alpha }\mathcal{D}%
\varphi _{\alpha }^{\ast }\exp \left\{ i\int d^{4}x\left( -\varphi ^{\ast
}D_{\mu }\varphi ^{\mu }+\varphi ^{\ast \mu }D_{\mu }\varphi \right. \right.
\notag \\
&&\left. \left. -m\left( \varphi ^{\ast }\varphi +\varphi ^{\ast \mu
}\varphi _{\mu }\right) \right) \right\}  \notag \\
&=&C\int \mathcal{D\varphi }^{\ast }\mathcal{D}\varphi \exp \left\{ -\frac{i%
}{m}\int d^{4}x\mathcal{\varphi }^{\ast }\left( D_{\mu }D^{\mu
}+m^{2}\right) \varphi \right\}  \notag \\
&=&\det G\left( x,y,A\right) =\exp Tr\ln G\left( x,x,A\right)  \notag
\end{eqnarray}

where 
\begin{equation}
G\left( x,y,A\right) =\left( D_{\mu }D^{\mu }+m^{2}\right) ^{-1}\delta
^{4}\left( x-y\right)  \label{ec6}
\end{equation}
is the GF of KGF equation in EM field $A_{\mu }(x)$. Thus utilizing Eqs. (%
\ref{ec3}), (\ref{ec5}), (\ref{ec6}) we get 
\begin{eqnarray}
Z\left( \overline{\mathcal{J}},\mathcal{J},\mathcal{J}_{\mu }\right)
&=&Z_{0}^{-1}\int \mathcal{D}A_{\mu }\exp \left\{ i\int d^{4}x\left( -\frac{1%
}{4}F_{\mu \nu }F^{\mu \nu }-\frac{1}{2\alpha }\left( \partial _{\mu }A^{\mu
}\right) ^{2}\right. \right.  \label{ec7} \\
&&\left. \left. +\mathcal{J}_{\mu }A^{\mu }+Tr\ln G\left( x,x,A\right) -\int
d^{4}y\overline{\mathcal{J}}\left( x\right) S\left( x,y,A\right) \mathcal{J}%
\left( y\right) \right) \right\}  \notag
\end{eqnarray}

From Eqs. (\ref{ec6}), (\ref{ec7}) we get to important conclusions: The all
many photons GF (Not only matrix elements of $S$-matrix for real photons )
coincide in DKP and KGF theories.

In general case one can obtain the generating functional or GF in external $%
A_{\mu }^{ex}$ by sustitution in Eqs. (\ref{ec1}), (\ref{ec2}) 
\begin{equation}
D^{\mu }\rightarrow \partial ^{\mu }-ie\left( A^{\mu }+A_{ex}^{\mu }\right)
\label{ec9}
\end{equation}

However, if $A_{ex}^{\mu }$ is not so strong to create pairs of particles,
the generating functional in $A_{ex}^{\mu }$ will be equal to: 
\begin{equation}
Z^{ex}=\exp \left\{ -i\int d^{4}xd^{4}y\overline{\mathcal{J}}\left( x\right)
S\left( x,y,A^{ex}\right) \mathcal{J}\left( y\right) \right\}  \label{ec10}
\end{equation}
and particles do not interact betweem themselves but only with $A_{ex}^{\mu
} $.

Now we consider interaction DKP spin-$0$ charged particles with external
gravitational fields. The Lagrangian density in this case can be written in
the form \cite{Lunardi} 
\begin{eqnarray}
\mathcal{L} &=&\sqrt{-g}\left\{ \frac{i}{2}\left[ \overline{\psi }\beta
^{\mu }\left( \partial _{\mu }+\frac{1}{2}\omega _{\mu }^{ab}S_{ab}\right)
\psi \right. \right.  \label{ec11} \\
&&\left. \left. -\left( \partial _{\mu }\overline{\psi }-\frac{1}{2}\omega
_{\mu }^{ab}\overline{\psi }S_{ab}\right) \beta ^{\mu }\psi -m\overline{\psi 
}\psi \right] \right\}  \notag
\end{eqnarray}

Here as in case Dirac equation, we have to introduce nontrivial tetrad field 
$e_{a}^{\mu }\left( x\right) $, which can be used to define the Riemannian
metric ($a$-vector Lorentz index; $\mu $-riemannian one): 
\begin{equation}
g_{\mu \nu }=e_{\mu }^{a}e_{\nu }^{b}\eta _{ab},\quad \eta _{ab}=diag\left\{
1,-1,-1,-1\right\}  \label{ec12}
\end{equation}
with properties: 
\begin{eqnarray}
e_{\mu }^{a}e_{a}^{\nu } &=&\delta _{\mu }^{\nu },\quad e_{\mu }^{a}e^{\mu
b}=\eta ^{ab}  \label{ec13} \\
\sqrt{-g} &=&\det e_{a}^{\mu }=e  \notag
\end{eqnarray}
$\beta _{\mu }=e_{\mu }^{a}\beta _{a}$, where $5\times 5$ matrix $\beta _{a}$
satisfy usual DKP commutation relations \cite{Kemmer}; the connection $%
\omega _{\mu }^{ab}$ equal to (see for instance the book \cite{Sabbata}): 
\begin{equation}
\omega _{\mu }^{ab}=e_{\lambda }^{a}e^{\nu b}\Gamma _{\mu \nu }^{\lambda
}-e^{\nu b}\partial _{\mu }e_{\nu }^{a}  \label{ec14}
\end{equation}

We consider two different cases: without torsion or with one. Without
torsion Levi-Civita connection is 
\begin{equation}
\Gamma _{\mu \nu }^{\lambda }=\Gamma _{\nu \mu }^{\lambda }=\overset{\circ }{%
\Gamma }_{\mu \nu }^{\lambda }=\frac{1}{2}\overset{\circ }{g}^{\lambda \rho
}\left( \partial _{\mu }\overset{\circ }{g}_{\rho \nu }+\partial _{\nu }%
\overset{\circ }{g}_{\rho \mu }-\partial _{\rho }\overset{\circ }{g}_{\mu
\nu }\right)  \label{ec15}
\end{equation}

and 
\begin{equation}
\partial _{\mu }\sqrt{-g}=\frac{\sqrt{-g}}{2}g^{\rho \lambda }\partial _{\mu
}g_{\rho \lambda }\equiv \sqrt{-g}\overset{o}{\Gamma }_{\rho \mu }^{\rho }
\label{ec17}
\end{equation}
If torsion is not equal zero $\Gamma _{\mu \nu }^{\lambda }$ can be defined
as Cartan connection, \cite{Andrade1}: 
\begin{equation}
\Gamma _{\mu \nu }^{\lambda }=e_{a}^{\lambda }\partial _{\nu }e_{\mu }^{a}
\label{ec18}
\end{equation}

Now from Eqs. (\ref{ec14}) and (\ref{ec18}) we obtain: 
\begin{equation}
\omega _{\mu }^{ab}=e_{\lambda }^{b}e^{\nu a}\left( \Gamma _{\mu \nu
}^{\lambda }-\Gamma _{\nu \mu }^{\lambda }\right)  \label{ec19}
\end{equation}
or 
\begin{equation}
\omega _{\mu \nu }^{\lambda }=e_{\nu a}e_{b}^{\lambda }\omega _{\mu
}^{ab}=\Gamma _{\mu \nu }^{\lambda }-\Gamma _{\nu \mu }^{\lambda }\equiv
2Q_{\mu \rho }^{\lambda }  \label{ec20}
\end{equation}
where the torsion $Q_{\mu \rho }^{\lambda }$ is a third rank tensor \cite
{Sabbata}. Thus we have two different Lagrangian depending on Eqs. (\ref
{ec15}) or (\ref{ec18}) which define the $\overset{o}{\omega }_{\mu }^{ab}$
or $\omega _{\mu }^{ab}$ in Eq. (\ref{ec19}). However, in both cases from
Eq. (\ref{ec11}) we get the form-equivalent equations for $\psi $ (The
equation for $\psi \left( x\right) $ in case Einstein-Cartan gravity \cite
{Sabbata} is discussed shortly in p.2 in Conclusion): 
\begin{equation}
\left( i\beta ^{\mu }\left( \partial _{\mu }+\frac{1}{2}\omega _{\mu
}^{ab}S_{ab}\right) -m\right) \psi \equiv \left( i\beta _{\mu }\nabla ^{\mu
}-m\right) \psi =0  \label{ec21}
\end{equation}
To show it in case (\ref{ec19}) when for instance torsion not equals zero we
can transform Eq. (\ref{ec11}) (omiting total derivatives) to the following
view: 
\begin{eqnarray}
\mathcal{L} &=&\sqrt{-g}\left\{ i\overline{\psi }\left( \beta _{\mu
}\partial ^{\mu }+\frac{1}{2}\omega _{\mu }^{ab}\beta ^{\mu }S_{ab}-m\right)
\psi \right.  \label{ec22} \\
&&\left. +\frac{i}{2}\overline{\psi }\left( \partial _{\mu }\left( \beta
^{\mu }\sqrt{-g}\right) -\omega _{\mu }^{\mu b}\beta _{b}\right) \psi
\right\}  \notag
\end{eqnarray}
We utilize at the derivation of Eq. (\ref{ec22}) the following equation: 
\begin{equation}
\omega _{\mu }^{ab}S_{ab}\beta ^{\mu }=\omega _{\mu }^{ab}\beta ^{\mu
}S_{ab}-2\omega _{\mu }^{\mu a}\beta _{a}  \label{23}
\end{equation}
One shows that the last term in Eq. (\ref{ec22}) equals zero: 
\begin{eqnarray}
\partial _{\mu }\left( \sqrt{-g}\beta ^{\mu }\right) &=&\beta ^{a}\partial
_{\mu }\left( e_{a}^{\mu }e\right) =\beta ^{a}\left[ \left( \partial _{\mu
}e_{a}^{\mu }\right) e+e_{a}^{\mu }\partial _{\mu }e\right]  \label{ec24} \\
&=&e\beta ^{a}\left[ \left( \partial _{\mu }e_{a}^{\mu }\right) +e^{\mu
}\Gamma _{\rho \mu }^{\rho }\right] =e\beta ^{a}\omega _{\mu
a}^{b}e_{b}^{\mu }=e\omega _{\mu }^{\mu b}\beta _{b}  \notag
\end{eqnarray}
where we used that \cite{Lunardi,Sabbata} $\left( \partial _{\mu }+\omega
_{\mu a}^{b}-\Gamma _{\lambda \mu }^{\lambda }\right) e_{a}^{\mu }=0$. Thus
the last term in Eq. (\ref{ec22}) equals zero. We get the same answer
without torsion, too. Then Lagrangian (\ref{ec22}) can be written so: 
\begin{equation}
\mathcal{L}=\sqrt{-g}\left\{ i\overline{\psi }\left( \beta _{\mu }\nabla
^{\mu }-m\right) \psi \right\}  \label{ec25}
\end{equation}
One writes down the Lagrangian (\ref{ec11}),(\ref{ec25}) in component form
utilizing the expressions for $\psi _{\alpha }(x),\overline{\psi }_{\alpha
}(x)$: 
\begin{eqnarray}
\psi _{\alpha } &=&\left( \varphi ,\varphi ^{0},\varphi ^{1},\varphi
^{2},\varphi ^{3},\right) \quad  \label{ec26} \\
\overline{\psi }_{\alpha } &=&\left( \psi ^{\ast }\eta \right) _{\alpha
}=\left( \varphi ^{\ast },\varphi ^{\ast 0},-\varphi ^{\ast 1},-\varphi
^{\ast 2},-\varphi ^{\ast 3},\right)  \notag
\end{eqnarray}
we obtain\footnote{%
For neutral particle $\mathcal{L}=\frac{1}{2}\sqrt{-g}\left\{ -\varphi
D_{\mu }\varphi ^{\mu }+\varphi ^{\mu }\partial _{\mu }\varphi -m^{2}\left(
\varphi ^{2}+\varphi _{\mu }\varphi ^{\mu }\right) \right\} $} 
\begin{equation}
L=\sqrt{-g}\left\{ -\varphi ^{\ast }D_{\mu }\varphi ^{\mu }+\varphi ^{\ast
\mu }\partial _{\mu }\varphi -m\left( \varphi ^{2}+\varphi _{\mu }^{\ast
}\varphi ^{\mu }\right) \right\}  \label{ec27}
\end{equation}
where 
\begin{equation}
D_{\mu }=\partial _{\mu }+\Gamma _{\nu \mu }^{\nu }  \label{ec28}
\end{equation}
From Eqs. (\ref{ec27}), (\ref{ec28}) we get the following $\mathcal{L}$%
-equations \cite{Andrade1}: 
\begin{eqnarray}
D_{\mu }\varphi ^{\mu }+m\varphi &=&0,\quad \partial _{\mu }\varphi
+m\varphi _{\mu }=0  \label{ec29} \\
D_{\mu }\partial ^{\mu }\varphi +m^{2}\varphi &=&0  \label{ec30}
\end{eqnarray}

The same equations arise for $\varphi ^{\ast }$.

If $D_{\mu }=\overset{\circ }{D}_{\mu }\equiv \partial _{\mu }+\overset{%
\circ }{\Gamma }_{\nu \mu }^{\nu }$ one gets KGF equation without torsion
(see Eqs. (\ref{ec20}), (\ref{ec21}) in \cite{Andrade1}); if $\Gamma _{\mu
\nu }^{\lambda }$ is defined by Eq. (\ref{ec18}) we get equation for scalar
particles with torsion (see (\ref{ec29}), (\ref{ec30}) in \cite{Andrade1}).

Thus on the classical level it is proved equivalence of DKP and KGF
equations in external gravitational fields \cite{Lunardi}.

On the quantum level ($2^{nd}$-quantization) we must construct (also as in
external $A^{ex}$) generating functional for GF of spin-$0$ particles,
interacting with external gravitational field $e_{\mu }^{a}\left( x\right) $%
. By definition the such generating functional in $\mathcal{L}$-approach is
(see Eq. (\ref{ec25})): 
\begin{equation}
Z\left( \overline{\mathcal{J}},\mathcal{J}\right) =Z_{0}^{-1}\int \mathcal{D}%
\psi \mathcal{D}\overline{\psi }\exp \left\{ i\int d^{4}x\sqrt{-g}\left( 
\overline{\psi }\left( i\beta _{\mu }\nabla ^{\mu }-m\right) \psi +\overline{%
\mathcal{J}}\psi +\overline{\psi }\mathcal{J}\right) \right\}  \label{ec31}
\end{equation}
where $Z_{0}=Z\left( 0,0\right) ,$ $\overline{\mathcal{J}}$, $\mathcal{J}$
are external currents.

Integrating in Eq. (\ref{ec31}) over $\overline{\psi }$ and $\psi $\ we
obtain: 
\begin{equation}
Z\left( \overline{\mathcal{J}},\mathcal{J}\right) =\exp \left\{ -i\int
d^{4}xd^{4}ye\left( x\right) e\left( y\right) \overline{\mathcal{J}}%
(x)S\left( x,y,e\right) \mathcal{J}\left( y\right) \right\}  \label{ec32}
\end{equation}

Here we have introduced the total GF of DKP particle in external
gravitational field $e_{a}^{\mu }$: 
\begin{equation}
S\left( x,y,e\right) =\left( i\beta _{\mu }\nabla _{x}^{\mu }-m\right)
^{-1}\delta ^{4}\left( x-y\right) e^{-1}\left( y\right)  \label{ec33}
\end{equation}

From Eq. (\ref{ec31}) we also have: 
\begin{equation}
S\left( x,y,e\right) =-i\left\langle 0\right| T\widehat{\psi }\left(
x\right) \widehat{\overline{\psi }}\left( y\right) \left| 0\right\rangle
\label{ec34}
\end{equation}

We suppose that external $e_{\mu }^{a}\left( x\right) $ is enough weak to
create pairs of new particles. Thus one can consider, in this case, only one
particle GF in external $e_{\mu }^{a}\left( x\right) $, since each particle
interacts with $e_{\mu }^{a}\left( x\right) $, but not between themselves.

\section{Equivalence between physical matrix elements of S-matrix in DKP and
KGF equations}

We use LSZ reduction formulas \cite{Lehmann} for proof of the equivalence of
the both theories. The main goal of LSZ approach is to express the matrix
elements of $S$-matrix through total many particles GF, for instance, Eq. (%
\ref{ec5}), making minimal assumptions as possible. To apply this approach
to DKP theory we must to write down the operatore's solution of the free DKP
equations. Taking into account that in DKP theory there are only two
linearly independent solutions \cite{Akhiezer} of the free equation ,one can
write: 
\begin{eqnarray}
\widehat{\psi }_{\substack{ in  \\ out}}\left( x\right) &=&\frac{1}{\left(
2\pi \right) ^{3/2}}\int d^{3}p\left\{ u^{-}\left( \mathbf{p}\right) 
\widehat{a}_{\substack{ in  \\ out}}^{-}\left( \mathbf{p}\right)
e^{-ipx}+u^{+}\left( \mathbf{p}\right) \widehat{b}_{\substack{ in  \\ out}}%
^{+}\left( \mathbf{p}\right) e^{ipx}\right\}  \label{ec35} \\
\widehat{\overline{\psi }}_{\substack{ in  \\ out}}\left( x\right) &=&\frac{1%
}{\left( 2\pi \right) ^{3/2}}\int d^{3}p\left\{ \overline{u}^{-}\left( 
\mathbf{p}\right) \widehat{a}_{\substack{ in  \\ out}}^{+}\left( \mathbf{p}%
\right) e^{ipx}+\overline{u}^{+}\left( \mathbf{p}\right) \widehat{b} 
_{\substack{ in  \\ out}}^{-}\left( \mathbf{p}\right) e^{-ipx}\right\}
\label{ec36}
\end{eqnarray}
here 
\begin{eqnarray}
\left( \widehat{p}\mp m\right) u^{\pm }\left( \mathbf{p}\right) &=&0,\quad 
\overline{u}^{\mp }\left( \mathbf{p}\right) \left( \widehat{p}\pm m\right)
=0,\quad \widehat{p}=\beta _{\mu }p^{\mu }  \label{ec37} \\
p_{0} &=&\left( \mathbf{p}^{2}+m^{2}\right) ^{1/2}\equiv \omega \left( 
\mathbf{p}\right) =\omega  \notag
\end{eqnarray}
operators $\widehat{a}_{\substack{ in  \\ out}}^{\pm },$ $\widehat{b} 
_{\substack{ in  \\ out}}^{\pm }$ satisfly to usual commutation relations
and the solutions in component from are: 
\begin{equation}
u_{\alpha }^{\pm }=\sqrt{\frac{m}{2\omega }}\left( 1,\pm \frac{i\omega }{m}%
,\pm \frac{ip^{1}}{m},\pm \frac{ip^{2}}{m},\pm \frac{ip^{3}}{m}\right)
\label{ec38}
\end{equation}

It is easy to check the scalar products are: 
\begin{eqnarray}
u^{\ast \mp }(\mathbf{p})\beta _{0}u^{\mp }(\mathbf{p}) &=&\overline{u}^{\mp
}(\mathbf{p})\beta _{0}u^{\mp }(\mathbf{p})=\pm 1  \label{ec39} \\
\overline{u}^{\pm }(\mathbf{p})\beta _{0}u^{\mp }(\mathbf{p}) &=&0  \notag
\end{eqnarray}

Operatores $\widehat{a}_{\substack{ in  \\ out}}^{\pm },$ $\widehat{b} 
_{\substack{ in  \\ out}}^{\pm }$ are that of creation and annihilation with
positive and negative charges accordingly.

From Eqs. (\ref{ec35}) and (\ref{ec36}) one also has: 
\begin{eqnarray}
\widehat{a}_{\substack{ in \\ out}}^{-}\left( \mathbf{p}\right)  &=&\frac{1}{%
\left( 2\pi \right) ^{3/2}}\int d^{3}xe^{-ipx}\overline{u}^{-}(\mathbf{p}%
)\beta _{0}\widehat{\psi }_{\substack{ in \\ out}}\left( x\right) 
\label{ec40} \\
\widehat{b}_{\substack{ in \\ out}}^{+}\left( \mathbf{p}\right)  &=&\frac{1}{%
\left( 2\pi \right) ^{3/2}}\int d^{3}xe^{ipx}\overline{u}^{+}(\mathbf{p}%
)\beta _{0}\widehat{\psi }_{\substack{ in \\ out}}\left( x\right) 
\label{ec41} \\
\widehat{a}_{\substack{ in \\ out}}^{+}\left( \mathbf{p}\right)  &=&\left( 
\widehat{a}_{\substack{ in \\ out}}^{-}\left( \mathbf{p}\right) \right)
^{\ast },\quad \widehat{b}_{\substack{ in \\ out}}^{-}\left( \mathbf{p}%
\right) =\left( \widehat{b}_{\substack{ in \\ out}}^{+}\left( \mathbf{p}%
\right) \right) ^{\ast }  \label{ec42}
\end{eqnarray}
However, decomposition (\ref{ec35}), (\ref{ec36}) of the operators $\widehat{%
\psi }_{\substack{ in \\ out}}\left( x\right) $, $\widehat{\overline{\psi }}
_{\substack{ in \\ out}}\left( x\right) $ over the operators $\widehat{a}
_{\substack{ in \\ out}}^{\pm }\left( \mathbf{p}\right) $, $\widehat{b}
_{\substack{ in \\ out}}^{\pm }\left( \mathbf{p}\right) $, which are that of
creation and annihilation of the plane wave do not give possibility to
construct the total set of the normalizale physical states in Hilbert space.
Moreover, utilizing these states for construction reduction formulas of LSZ
we can not carry out strictly matematically proof of the equivalence of DKP
and KGF theories. For these aims we have to construct new operators which
create and annihilate the wave packets states. We define the new operators
in the following decomposition (we omit sign \symbol{94} over operators) 
\begin{eqnarray}
a_{\substack{ in \\ out}}^{\mp }\left( \mathbf{p}\right) 
&=&\sum_{n=0}^{\infty }f_{n}\left( \mathbf{p}\right) a_{\substack{ in \\ out
}}^{\mp }\left( \mathbf{n}\right)   \label{ec43} \\
b_{\substack{ in \\ out}}^{\mp }\left( \mathbf{p}\right) 
&=&\sum_{n=0}^{\infty }f_{n}\left( \mathbf{p}\right) b_{\substack{ in \\ out
}}^{\mp }\left( \mathbf{n}\right)   \notag
\end{eqnarray}
Here $f_{n}\left( \mathbf{p}\right) =$ $f_{n}^{\ast }\left( \mathbf{p}%
\right) $ are total ortonormalized set of functions with properties\footnote{%
We choose $f_{n}\left( \mathbf{p}\right) =f_{n}^{\ast }\left( \mathbf{p}%
\right) =\exp \left( -\frac{\mathbf{p}^{2}}{2m^{2}}\right) \Pi
_{i=1}^{3}H_{n}\left( \frac{p_{i}}{m}\right) $, where $H_{n}\left( x\right) $%
\ denotes a Hermite polinomial of degree $n$.}  
\begin{equation}
\int f_{n}\left( \mathbf{p}\right) f_{m}\left( \mathbf{p}\right) d^{3}%
\mathbf{p=\delta }_{nm},\quad \sum_{n=0}^{\infty }f_{n}\left( \mathbf{p}%
\right) f_{n}\left( \mathbf{q}\right) \mathbf{=\delta }^{3}\left( \mathbf{p}-%
\mathbf{q}\right)   \label{ec44}
\end{equation}
from Eqs.(\ref{ec43}) and (\ref{ec44}) we get: 
\begin{equation}
a_{\substack{ in \\ out}}^{\pm }\left( \mathbf{n}\right) =\int d^{3}\mathbf{p%
}f_{n}\left( \mathbf{p}\right) a_{\substack{ in \\ out}}^{\pm }\left( 
\mathbf{p}\right) \mathbf{,\quad }b_{\substack{ in \\ out}}^{\pm }\left( 
\mathbf{n}\right) =\int d^{3}\mathbf{p}f_{n}\left( \mathbf{p}\right) b
_{\substack{ in \\ out}}^{\pm }\left( \mathbf{p}\right)   \label{ec45}
\end{equation}
From commutations relations: 
\begin{equation}
\left[ a_{\substack{ in \\ out}}^{-}\left( \mathbf{p}\right) ,a_{\substack{ %
in \\ out}}^{+}\left( \mathbf{q}\right) \right] =\left[ b_{\substack{ in \\ %
out}}^{+}\left( \mathbf{p}\right) ,b_{\substack{ in \\ out}}^{+}\left( 
\mathbf{q}\right) \right] =\delta ^{3}\left( \mathbf{p}-\mathbf{q}\right) 
\label{ec46}
\end{equation}
and from Eqs. (\ref{ec43})-(\ref{ec45}) one get: 
\begin{equation}
\left[ a_{\substack{ in \\ out}}^{-}\left( n\right) ,a_{\substack{ in \\ out
}}^{+}\left( m\right) \right] =\left[ b_{\substack{ in \\ out}}^{-}\left(
n\right) ,b_{\substack{ in \\ out}}^{+}\left( m\right) \right] =\delta _{mn}
\label{c47}
\end{equation}
Vaccuum and $S$-matrix are defined as usual 
\begin{eqnarray}
a_{\substack{ in \\ out}}^{-}\left( n\right) \left| 0\right\rangle 
_{\substack{ in \\ out}} &=&b_{\substack{ in \\ out}}^{-}\left( n\right)
\left| 0\right\rangle _{\substack{ in \\ out}}=0,\quad S\left|
0\right\rangle _{\substack{ in \\ out}}=\left| 0\right\rangle _{\substack{ in
\\ out}}  \label{ec48} \\
a_{out}^{+} &=&a_{in}^{+}S  \label{ec49}
\end{eqnarray}
The any physical matrix element of $S$-matrix one has: 
\begin{equation}
\left\langle n^{\prime },m^{\prime };out\right. \left| n,m;in\right\rangle
=\left\langle n^{\prime },m^{\prime };in\right| S\left| n,m;in\right\rangle 
\label{ec50}
\end{equation}
were $n+m=n^{\prime }+m^{\prime }$ is conservation of charge and 
\begin{equation}
\left| n,m;_{out}^{in}\right\rangle \equiv \overset{n}{\underset{i=1}{\Pi }}%
\overset{m}{\underset{j=1}{\Pi }}a_{\substack{ in \\ out}}^{+}\left(
n_{i}\right) b_{\substack{ in \\ out}}^{+}\left( n_{i}\right) \left|
0\right\rangle   \label{ec51}
\end{equation}

Now we can formulate main assumption of LSZ approach\footnote{%
This assumption can be proved in case of microcasuality theories, when
comutator $\left[ \psi \left( x\right) ,\overline{\psi }\left( y\right) %
\right] =0,\quad \left( x-y\right) ^{2}<0$}: for any matrix elemnts of
Heisenberg operators $\widehat{\psi }\left( x\right) $\ and $\widehat{%
\overline{\psi }}\left( x\right) $ the following asymptoptic relation are
implemented: 
\begin{equation}
\lim_{x_{0}=\mp \infty }\left\langle n^{\prime },m^{\prime
};_{out}^{in}\right| \widehat{\psi }\left( x\right) \left|
n,m;_{out}^{in}\right\rangle =\left\langle n^{\prime },m^{\prime
};_{out}^{in}\right| \widehat{\psi }_{\substack{ in  \\ out}}\left( x\right)
\left| n,m;_{out}^{in}\right\rangle  \label{ec52}
\end{equation}
and the same relations for $\widehat{\overline{\psi }}\left( x\right) $.

In order to do not complicate the proof of the equivalence of the DKP and
KGF theories we restricted by considering of the matrix elements of $S$%
-matrix for particles with the same (positive) charges and one utilizes LSZ
reduction formula. We have: 
\begin{eqnarray}
\left\langle 0\right| \overset{k}{\underset{i=1}{\Pi }}a_{out}^{-}\left(
n_{i}\right) \overset{k}{\underset{j=1}{\Pi }}a_{in}^{+}\left( m_{j}\right)
\left| 0\right\rangle  &=&\left\langle 0\right| \overset{k-1}{\underset{i=1}{%
\Pi }}a_{out}^{-}\left( n_{i}\right) a_{out}^{-}\left( n_{k}\right) \overset{%
k}{\underset{j=1}{\Pi }}a_{in}^{+}\left( m_{j}\right) \left| 0\right\rangle 
\notag \\
&=&C\left\langle n_{1},...,n_{k-1};out\right| \int d^{4}xf_{n_{k}}^{\ast
-}\left( x\right)   \notag \\
&&\left( i\beta _{\mu }\overrightarrow{\partial }_{x}^{\mu }-m\right) 
\widehat{\psi }\left( x\right) \left| n_{1},...,n_{k};in\right\rangle 
\label{ec53}
\end{eqnarray}
Here 
\begin{equation}
f_{\alpha ;n_{k}}^{\ast -}\left( x\right) =\frac{1}{\left( 2\pi \right)
^{3/2}}\int d^{3}p\overline{u}_{\alpha }^{-}\left( \mathbf{p}\right)
f_{n_{k}}\left( \mathbf{p}\right) e^{ipx},\quad p_{0}=\omega \left( \mathbf{p%
}\right)   \label{ec54}
\end{equation}
and $C$- not essential constant

Utilizing Eqs. (\ref{ec26}), (\ref{ec38}), (\ref{ec54}) we can rewrite Eq. (%
\ref{ec53}) in component form: 
\begin{eqnarray}
&&\left\langle n_{1},...,n_{k-1};out\right| \int d^{4}xd^{3}\mathbf{p}%
f_{n_{k}}\left( \mathbf{p}\right) \left\{ -\frac{1}{m}e^{ipx}\left( 
\overrightarrow{\square }_{x}+m^{2}\right) \widehat{\varphi }\left( x\right)
\right.  \label{ec55} \\
&&\left. +\frac{\partial }{\partial x^{\mu }}\left[ e^{ipx}\left( \frac{1}{m}%
\partial _{x}^{\mu }\widehat{\varphi }\left( x\right) -\widehat{\varphi }%
^{\mu }\left( x\right) \right) \right] \right\} \left|
n_{1},...,n_{k};in\right\rangle  \notag
\end{eqnarray}

The main idea of the proof is to show that the second term under total
derivative in Eq. (\ref{ec55}) is equal to zero. We have\footnote{%
It's easy to show that term in (\ref{ec56}) under total derivative has no $%
\delta $-function singularities.} 
\begin{eqnarray}
&&\int d^{4}x\int d^{3}\mathbf{p}f_{n_{k}}\left( \mathbf{p}\right) \frac{%
\partial }{\partial x^{\mu }}\left[ e^{ipx}\left( \frac{1}{m}\partial
_{x}^{\mu }\widehat{\varphi }\left( x\right) -\widehat{\varphi }^{\mu
}\left( x\right) \right) \right]  \label{ec56} \\
&=&\int d\sigma _{\mu }\int d^{3}\mathbf{p}f_{n_{k}}\left( \mathbf{p}\right)
e^{ipx}\left( \frac{1}{m}\partial _{x}^{\mu }\widehat{\varphi }\left(
x\right) -\widehat{\varphi }^{\mu }\left( x\right) \right)  \notag
\end{eqnarray}

One chooses the surface $\sigma _{\mu }$ so: 
\begin{equation}
\sigma _{\mu }:\left\{ -T\leq x_{0}\leq T;\quad -L\leq x_{i}\leq L;\quad
i=1,2,3\right\}  \label{ec57}
\end{equation}

The firs term, $\mu =0$ equals 
\begin{eqnarray}
&&\int d^{3}x\left[ \int d^{3}\mathbf{p}f_{n_{k}}\left( \mathbf{p}\right)
e^{i\omega T+i\mathbf{px}}\left( \frac{1}{m}\frac{\partial }{\partial T}%
\widehat{\varphi }\left( \mathbf{x},T\right) -\widehat{\varphi }^{0}\left( 
\mathbf{x},T\right) \right) \right.  \label{ec58} \\
&&\left. -\int d^{3}\mathbf{p}f_{n_{k}}\left( \mathbf{p}\right) e^{-i\omega
T+i\mathbf{px}}\left( \frac{1}{m}\frac{\partial }{\partial T}\widehat{%
\varphi }\left( \mathbf{x},-T\right) -\widehat{\varphi }^{0}\left( \mathbf{x}%
,-T\right) \right) \right]  \notag
\end{eqnarray}

Since 
\begin{equation}
\lim_{T\rightarrow \infty }\frac{\partial }{\partial T}\widehat{\varphi }%
\left( \mathbf{x},\pm T\right) =\frac{\partial }{\partial T}\widehat{\varphi 
}_{\substack{ in  \\ out}}\left( \mathbf{x},\pm T\right) =\widehat{\varphi } 
_{\substack{ in  \\ out}}^{0}\left( \mathbf{x},\pm T\right)  \label{ec59}
\end{equation}
the first term (\ref{ec58}) disappears in limit $T\rightarrow \pm \infty $.

It is enough to consider only one term in Eqs. (\ref{ec56}) and (\ref{ec57}%
), for instance $i=1$, and to show that it goes to zero at $L\rightarrow \pm
\infty $ due to Gaussian properties of the packets $f_{n}\left( \mathbf{p}%
\right) $ (see$^{5}$ p.8)

The term $\mu =1$ is: 
\begin{eqnarray}
&&I(L)=\int dx_{0}dx_{\bot }\int d^{3}\mathbf{p}f\left( \mathbf{p}\right)
e^{i\omega x_{0}-ip_{\bot }x_{\bot }}\left\{ e^{-ip_{1}x_{1}}\left( \frac{1}{%
m}\partial _{x}^{1}\widehat{\varphi }\left( x\right) -\widehat{\varphi }%
^{1}\left( x\right) \right) _{x_{1}=L}\right.  \notag \\
&&\left. -e^{-ip_{1}x_{1}}\left( \frac{1}{m}\partial _{x}^{1}\widehat{%
\varphi }\left( x\right) -\widehat{\varphi }^{1}\left( x\right) \right)
_{x_{1}=-L}\right\}  \label{ec60}
\end{eqnarray}
where $p_{i}=\left( p_{1},p_{\bot }\right) ,$ $x_{i}=\left( x_{1},x_{\bot
}\right) $, $f_{n}=f_{n_{k}}$

One estimates the first term $\left( x_{1}=L\right) $ in Eq. (\ref{ec60}),
which can be writen in the form: 
\begin{eqnarray}
I\left( L\right) &\equiv &\int dp_{1}e^{-ip_{1}L}e^{-\frac{p_{1}^{2}}{2m^{2}}%
}H_{n}\left( \frac{p_{1}}{m}\right) \int dx_{0}dx_{\bot }\int d\mathbf{p}%
_{\bot }f_{n}\left( \mathbf{p}_{\bot }\right) e^{i\omega x_{0}-ip_{\bot
}x_{\bot }}  \notag \\
&&\left( \frac{1}{m}\partial _{x}^{1}\widehat{\varphi }\left( x\right) -%
\widehat{\varphi }^{1}\left( x\right) \right)  \label{ec61}
\end{eqnarray}

By definition the integral over $x_{0},x_{\bot },\mathbf{p}_{\bot }$ in Eq. (%
\ref{ec61}) is a distribution (generalized function) of polinomial grows or
Shwartz distribution, (renormalizable theory), i.e. it can grow not faster
than polinomial: $\leq C_{1}p_{1}^{l}|L|^{k}$, where $l$ and $k$ are some
numbers, $C_{1}$ is a constant.

Then 
\begin{eqnarray}
I\left( L\right) &\leq &C_{1}|L|^{k}\int dp_{1}p_{1}^{l}e^{-ip_{1}L}e^{-%
\frac{p_{1}^{2}}{2m^{2}}}H_{n}\left( \frac{p_{1}}{m}\right)  \label{ec62} \\
&<&C_{1}|L|^{k}\left| \left( -i\frac{\partial }{\partial L}\right)
^{l+n}\right| \int dp_{1}e^{-ip_{1}L-\frac{p_{1}^{2}}{2m^{2}}}  \notag \\
&=&C_{1}|L|^{k}\left| \left( -i\frac{\partial }{\partial L}\right)
^{l+n}\right| e^{-\left( Lm\right) ^{2}}  \notag
\end{eqnarray}

Thus $I\left( L\right) $ decreases as $\sim e^{-\left( Lm\right)
^{2}}|L|^{k+l+n}$, and we proved that the second term in Eq. (\ref{ec55})
equals zero.

If we repeat the LSZ procedure for the second operator $a_{out}^{+}\left(
n\right) $ in Eq. (\ref{ec53}) we get 
\begin{eqnarray}
\left\langle k;out\right| \left. k;in\right\rangle &=&C\left\langle
k-2;out\right| \int d^{4}x_{1}d^{4}x_{2}d^{3}\mathbf{p}_{1}d^{3}\mathbf{p}%
_{2}e^{ip_{1}x_{1}}f_{n_{1}}^{\ast }\left( \mathbf{p}_{1}\right)
f_{n_{2}}\left( \mathbf{p}_{2}\right)  \notag \\
&&\left( \frac{\overrightarrow{\square }_{1}+m^{2}}{m}\right) \left\{
e^{ip_{2}x_{2}}\left( \frac{\overrightarrow{\square }_{2}+m^{2}}{m}\right) T%
\widehat{\varphi }\left( x_{1}\right) \widehat{\varphi }\left( x_{2}\right)
\right.  \notag \\
&&\left. +\partial _{\mu }^{x_{2}}\left[ e^{ip_{2}x_{2}}\left( \partial
_{x_{2}}^{\mu }T\widehat{\varphi }\left( x_{1}\right) \widehat{\varphi }%
\left( x_{2}\right) -T\widehat{\varphi }\left( x_{1}\right) \widehat{\varphi 
}\left( x_{2}\right) \right) \right] \right\} \left| k;in\right\rangle 
\notag \\
&&  \label{ec63}
\end{eqnarray}

Agian one can prove the last term under total derivative equals zero due to
properties of packets $f_{n}(\mathbf{q})$. Continuating this inductive
procedure we go to conclusion that all physical matrix elements of $S$%
-matrix in DKP theory coincide with that of in KGF theory independently of
character interaction on these theories, if in the both theories the LSZ
asymptotic conditions (\ref{ec52}) are implemented for Heisenberg operators%
\footnote{%
The quasilocal terms do not appear in Eq. (\ref{ec63}) \cite{Fainberg}.}.

From general result (\ref{ec63}) the equivalence between DKP and KGF
theories also follows in case interaction spin-$0$ particles with external
eletromagnetic field, $A_{\mu }^{ex}\left( x\right) $, and gravitational
fields, $e_{\mu a}^{ex}\left( x\right) $, see Eqs. (\ref{ec10}) and (\ref
{ec32})-(\ref{ec34}).

Now we briefly describe the proof of equivalence of the theories in case
interaction spin-$0$ particles with non-abelian (Yang-Mills) external and
quantized fields $A_{\mu }^{i}$, where $i=1,...,2N^{2}-1$, is the group
index of $SU(N)$ group.

One is restricted by case when spin-o particles in DKP and KGF theories are
taken on fundamental representation of the $SU(N)$ group.

The initial density of Lagrangian in DKP theory is (in $\alpha $-gauge) 
\begin{eqnarray}
\mathcal{L}_{DKP} &=&\overline{\psi }_{\alpha }\left( x\right) \left( i\beta
_{\mu }D_{ab}^{\mu }-m\delta _{ab}\right) \psi _{b}-\frac{1}{4}F_{\mu \nu
}^{i}F_{i}^{\mu \nu }-\frac{1}{2\alpha }\left( \partial _{\mu }A_{i}^{\mu
}\right) ^{2}  \label{ec64} \\
&&-\overline{C}_{a}\left( \partial _{\mu }D_{ab}^{\mu }C_{b}\right)  \notag
\end{eqnarray}

Here: $\left( D^{\mu }\right) _{ab}=\partial ^{\mu }\delta _{ab}-ig\left(
A_{i}^{\mu }T_{i}\right) _{ab},$ $a=1,...N$, index of fundamental
representation. $F_{\mu \nu }^{i}=\partial _{\mu }A_{\nu }^{i}-\partial
_{\nu }A_{\mu }^{i}+igf^{ijk}A_{\mu }^{j}A_{\nu }^{k};$ $A_{\mu }^{i}$-is
Yang-Mills field; $T_{i}$ are $N\times N$ matrixes with the commutation
relation 
\begin{equation}
\left[ T_{i},T_{j}\right] =if_{ijk}T_{k}  \label{ec65}
\end{equation}
$f_{ijk}$ are structure constant of $SU(N)$ group; $\overline{C}_{a},C_{b}$
Faddeev-Popov anticommuting ghost fields. We also write down only spin-$0$
part of \ $\mathcal{L}_{int}$ in KGF theory: 
\begin{equation}
\mathcal{L}_{KGF}=-\varphi _{a}\left[ \left( D_{\mu }\right) _{ab}\left(
D^{\mu }\right) _{bc}+m^{2}\delta _{ab}\right] \varphi _{c}  \label{ec66}
\end{equation}

Now, utilizing LSZ reduction formulas we shall obtain the same formulas as
Eqs. (\ref{ec55})-(\ref{ec63}) excluding appearance additional group index $%
a $ at all operators and at solution of free equations, $f_{n_{k}}^{\ast }$,
in Eqs. (\ref{ec53}) and (\ref{ec54}). Therefore, the proof of equivalence
of the physical matrix elements of $S$-matrix for spin-$0$ particles and for
GF of any number of the Yang-Milles particles can be carried out with help
of the same arguments as before.

\section{Conclusion}

\begin{enumerate}
\item  Starting from Lagrangian approach to DKP theory of spin-$0$ particles
we constructed generating functional for many particles GF of the theory
(see Eqs. (\ref{ec1}), (\ref{ec3}), (\ref{ec31})) and utilizing the LSZ
reduction formulas, we strictly proved the equivalence between physical
matrix elements of $S$-matrix in DKP and KGF theories, being the general
formula (\ref{ec63}) can be applied to any type of interaction on the both
theories, and consequantly to interaction spin-$0$ particles with quantized
Maxwell, Yang Milles fields and with external gravitational fields.We also
proved (see for instance Eqs. (\ref{ec5})-(\ref{ec7})) that in DKP and KGF
theories the many particles GF of photons and Yang-Mills particles exactly
coincide (not only physical matrix elements).

\item  Considering the interaction of DKP spin-$0$ particles with external
gravitational field we restricted by two cases: without torsion, when
connection $\Gamma _{\mu \nu }^{\lambda }=\overset{\circ }{\Gamma }_{\mu \nu
}^{\lambda }$(see Eq. (\ref{ec15})) is Levi-Civita one and with torsion,
when the total connection $\Gamma _{\mu \nu }^{\lambda }$ is Cartan
connection\footnote{%
One notes that in this case the ''metric postulate'' $\nabla _{\mu }g_{\nu
\lambda }=0$ does not implement.}, Eq. (\ref{ec18}), antisymmetrical part of
which, Eq. (\ref{ec20}), expressed through torsion tensor. This case is one
of teleparalled description of gravity \cite{Hayashi, Andrade, Maluf,
Andrade1, Andrade2}. We would like to note that there is the third case
interaction with gravitational field, which leads to Einstein-Cartan gravity 
\cite{Sabbata}. In the last case the metric postulate is implemented: 
\begin{equation}
\nabla _{\mu }g_{\nu \lambda }=\partial _{\mu }g_{\nu \lambda }-\Gamma _{\mu
\nu }^{\rho }g_{\rho \lambda }-\Gamma _{\mu \lambda }^{\rho }g_{\nu \rho }=0
\label{ec67}
\end{equation}
here $\Gamma _{\mu \nu }^{\rho }=\overset{\circ }{\Gamma }_{\mu \nu }^{\rho
}+^{\left( -\right) }\Gamma _{\mu \nu }^{\rho }$

where $^{\left( -\right) }\Gamma _{\mu \nu }^{\lambda }$ is antisymmetrical
part of Einstein-Cartan connection. The equation for $\psi $ now has the
view (instead(\ref{ec21})): 
\begin{equation}
\left( i\beta ^{\mu }\left( \nabla _{\mu }+^{\left( -\right) }\Gamma _{\mu
\rho }^{\rho }\right) -m\right) \psi =0  \label{ec68}
\end{equation}

The proof of the equivalence of DKP and KGF equations is the same.

\item  For unstable particles we suggest the following fenomenological
Lagrangian of interaction: 
\begin{equation}
\mathcal{L}_{int}=\lambda \left( \overline{\psi }^{M}\beta _{\mu }\psi ^{m}+%
\overline{\psi }^{m}\beta _{\mu }\psi ^{M}\right) j^{\mu }\left( x\right) 
\label{ec69}
\end{equation}

where $\psi ^{M},$ $\psi ^{m}$ are the DKP functions, decribing accordingly $%
K_{L}^{0}$ and $\pi $ mesons with masses $M$ and $m$; $j_{\mu }=\left( 
\overline{e}\gamma _{5}\gamma _{\mu }\nu \right) +\left( \overline{\nu }%
\gamma _{5}\gamma _{\mu }e\right) $; $\lambda $-fenomenological constant of
point-like interaction.

Then it is easy to show in $\lambda ^{2}$-approximation that the imaginary
part of GF of $K_{L}^{0}$ meson, which determines the amplitude of
probability decay $K_{L}^{0}\rightarrow \pi ^{+}+e^{-}+\overline{\nu }$, in
the DKP theory, concides with that of in KGF theories.
\end{enumerate}

\section{Acknowledgments}

V. Ya. F. thanks to FAPESP for support (grant number 98/06237-0) and RFFI
for partial support (grant number 99-01-00376). B. M. P. thanks to CNPq for
partial support.

\end{document}